# User requirements for inclusive technology for older adults

Mlađan Jovanović

*Department for Computer Science, Singidunum University, Belgrade, Serbia*

Antonella De Angeli

*Department for Computer Science, Free University of Bolzano, Bolzano, Italy*

Andrew McNeill

*Department of Psychology, Northumbria University, Newcastle Upon Tyne, UK*

Lynne Coventry

*Department of Psychology, Northumbria University, Newcastle Upon Tyne, UK*

mjovanovic@singidunum.ac.rs (corresponding author)

antonella.deangeli@unibz.it

andrew.mcneill@northumbria.ac.uk

lynne.coventry@northumbria.ac.uk

# User requirements for inclusive technology for older adults

**Abstract.** Active aging technologies are increasingly designed to support an active lifestyle. However, the way in which they are designed can raise different barriers to acceptance of and use by older adults. Their designers can adopt a negative stereotype of aging. Thorough understanding of user requirements is central to this problem. This paper investigates user requirements for technologies that encourage an active lifestyle and provide older people with the means to self-manage their physical, mental, and emotional health. This requires consideration of the person and the sociotechnical context of use. We describe our work in collecting and analyzing older adults' requirements for a technology which enables an active lifestyle. The main contribution of the paper is a model of user requirements for inclusive technology for older people.

## 1. Introduction [1]

An active lifestyle in later life is essential for older adults to maintain and facilitate their well-being. In our work, we adopt the World Health Organization's (WHO's) standard of age 60 and above as discriminant to consider a person as "older" (WHO, 2002). Previous research in Human-Computer Interaction (HCI) has been mainly focused on the physical decline in old age, defining it as a deficit to be solved by technology (Cozza et al., 2017; Pu et al., 2019; Yusif and Hafeez-Baig, 2016). Old age is described as multiple health infirmities and conditions that need to be taken care of (Vines et al., 2015b). Numerous technological interventions are developed to support healthcare and disability-free living in later life (Cozza et al., 2017; Peruzzini and Germani, 2014).

---

However, these technologies did not gain a larger momentum in practice due to a mismatch between the users' real needs and how the technology responds to these needs, instead treating them simply as frail people (Lee and Coughlin, 2015; Wilkinson and De Angeli, 2014). Therefore, it is crucial to collect older users' needs early in the design process and interpret them correctly.

Some researchers have been approaching older adults as users who actively manage their health and well-being (Carroll et al., 2012; Light et al., 2015; Light et al., 2016; Suwa et al., 2020; Vines et al., 2015a; Yuan et al., 2018). Acknowledging the desire for active engagement in later life, some research has investigated the role of physical activity and social interaction in later life. Physical activity is related to positive outcomes throughout the lifespan including for older adults. For example, increased physical activity reduced the risk of depression (Strawbridge et al., 2002), improved cognitive performance (Colcombe and Kramer, 2003) and improved quality of life (Sun et al., 2013). Social engagement is another critical factor for older adults' well-being. Social isolation poses significant risks to older adults mental health, resulting in loneliness and depression (Cornwell and Waite, 2009). Older people's perception of active aging has been primarily concerned with physical activity (Stenner et al., 2011). However, older people report that physical activity is entangled with social and mental activity, and does not exist in isolation (Gerling et al., 2017; Lazar and Nguyen, 2017; Litt et al., 2002; Vines et al., 2015a). This illustrates the complexities of the design space. Aging poses specific challenges to the physical body. However, the body exists within a social and material environment, which is still unaccounted for by Information and Communication Technology (ICT) research for older people.

Despite previous studies having addressed acceptance and adoption (Khosla et al., 2017; Macedo et al., 2017; Pu et al., 2019; Vassli et al., 2018), less is known on designing for older people while ensuring they are engaged from the beginning and helping them to reason about future possibilities. This paper provides a model of user requirements for active aging technology generated from two interview studies which occurred at different phases within our research (described in Section 4). We build on the model of user desires (De Angeli et al., 2020), which describes intentions to engage in behaviors in later life as desires for activities as follows: *feeling good* and *doing good* as attitudinal drivers; *autonomy* and *belonging* as social drivers; and *resilience* as a material driver.

The paper continues as follows. Section 2 introduces the ACANTO research project. Section 3 describes existing work on user requirements for aging technologies, gives an overview of the ICT for later life, and presents the model of desires derived from our initial study that focused on the older adults' preferred activities and the behavioral drivers to engage in the activities. The model informed the design of 14 scenarios illustrating use of the ACANTO system that are described in Section 4. In Section 5 we describe and report on the results of a user study (N=21) designed to evaluate the potential functionality of the ACANTO system by analyzing responses to a set of personas and scenarios. These scenarios were illustrated in an animation which illustrated different functions for the ACANTO system. Section 6 presents the derived model of user requirements. We discuss the results in Section 7 with design implications for future, more inclusive ICT requirement gathering and research for active aging. Section 8 concludes the paper.

## 2. The ACANTO project

The ACANTO project aimed to develop a sociotechnical system to maintain well-being in later life while facilitating physical activity and social interaction (ACANTO, 2018). The philosophy behind the ACANTO system design emphasized the need to create a positive experience for older people presenting the system as an opportunity rather than a necessity. The project had two underlying goals: 1) promoting well-being and health by encouraging mobility and social interaction, and 2) supporting rehabilitation and recovery monitoring after a fall. To achieve these goals, the design requires three technical components – a robotic walker, a social network and a recommender system. The components work independently or in connection depending on the specific needs of the user.

The walker, named FriWalk (Friendly Walker), aims to support the self-management of physical activities by older people, who may require a physical support to maintain independent mobility or facilitate rehabilitation. In addition, the walker can monitor of numerous physical attributes, including balance, strength, gait and walking speed. By acting as an assistant in physical exercises to maintain strength, balance, flexibility, or endurance, the walker can reduce the risk of falling (Danielsen et al., 2016). This is a common issue in later life, as each year about a third of older people that live independently experience a fall (Gillespie et al., 2012).

Prevention is a public health priority in this context as even a minor fall may require hospitalization. Older adults have a higher risk of mortality resulting from minor falls (Spaniolas et al., 2010). For those who survive, hospitalization can accelerate cognitive decline (Wilson et al., 2010). Falls even if not requiring hospitalization are associated with functional decline, reduced social and physical activities (Tinetti et al., 1998) and depression (Stel et al., 2004) and many never regain the level of mobility and confidence they enjoyed before falling (Schutzer and Graves, 2004).

The social network collects a variety of information about the older adults and creates their profiles. The social network links people with similar backgrounds and/or interests together. The recommender system uses the profiles to suggest new people to meet, activities to do together and places to visit by matching users based on a variety of factors, such as shared interests, background, location, and life circumstances. This way the recommender system can also support awareness of nearby, appropriate opportunities for activity and socializing. The social network and recommendation systems are accessed through a tablet which can be used with the walker or independently (Figure 1). We named the tablet FriTab (Friendly Tablet). These profiles are adapted as the older adults use the different parts of the ACANTO system and their mobility is assessed by the walker, they gradually provide information to the social network, or they rate the different activities they are recommended.

Older adults were involved at the beginning of the design process to help us understand the activities they like to engage in and how the ACANTO system may support them. This activity was facilitated by using the Integrated Behavioural Model (IBM) (Montano and Kasprzyk, 2008), which provided a structured framework and guided the interview to identify pragmatic as well as hedonic factors that make an activity desirable for older adults. In previous work, we named these factors as desires for activities. For example, proximity to people and activities was identified as an important pragmatic factor (De Angeli et al., 2020).

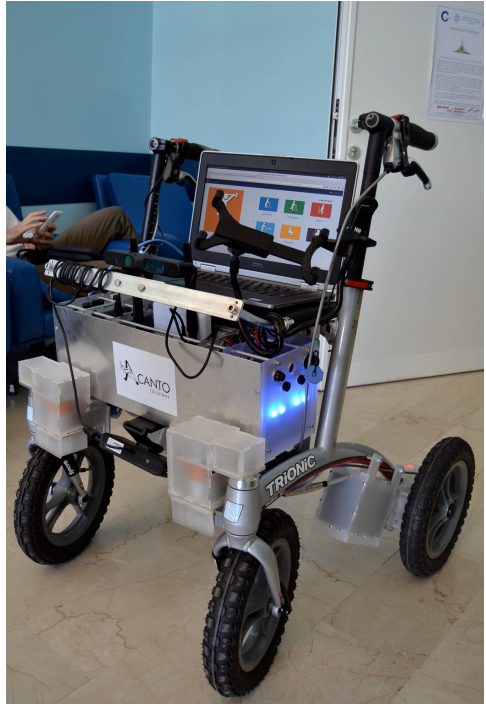

Figure 1. The ACANTO system - the robotic walker (FriWalk) with the mounted tablet (FriTab).

## 3. Related Work

This section starts with the existing work concerning older users' requirements for more inclusive technology. We continue with a literature review challenging mainstream assumptions of technologies for the older user. Finally, we elaborate on the model of desires used to drive user requirements elicitation for older user technology.

### *3.1. Older users' requirements for ICT*

Existing research concerning older users' requirements for technology, either seeks to validate theoretical models in practice, or are empirical studies of older users of technology. A number of different theoretical models can be applied to this context. For example, the Technology Acceptance Model (TAM) proposes perceived ease of use and perceived usefulness as the most influential predictors of technology acceptance (Czaja et al., 2006; Davis, 1989). The Unified Theory of Acceptance and Use of Technology (UTAUT) (Venkatesh et al., 2003) combines several theoretical models of acceptance, including the Theory of Planned Behaviour (TPM) (Ajzen, 1991). According to the UTAUT, four components predict the intention to use a technology: performance expectancy, effort expectancy, social influence and facilitating conditions. Research commonly does not use all constructs within a model, rather it looks at a subset of these. For example, Barnard et al. (2013) measured the effect of specific UTAUT's constructs - effort expectancy and facilitating conditions while assuming while assuming that performance expectancy and social influence constructs are met. However, performance expectancy and facilitating conditions appeared two most influential factors of older users' intention to use technology (Macedo, 2017). Another adaptation of the UTAUT model is the Almere model (Heerink et al., 2010), which has been used to understand the acceptance of social robots as a technology that has socially interactive functionality and in addition to the functional evaluation, it measures factors related to social interaction with rather than through the technology, enjoyment and trust. The Domestication Theory (Silverstone and Haddon, 1996) describes the steps that technology goes through to be accepted in a social setting. An example of this is the role of video games (De Schutter et al., 2015) where initial excitement was replaced with moral panic relating to potential to increase violence, to acceptance of benefits that can be brought and acceptance into society. De Schutter et al. (2015) highlighted that the meanings older adults attach to games originate from their lived experiences and daily routines. A similar process could be used to examine the role of social robots in society, where initial excitement was taken over by fear of non-human interaction for older adults, but the benefits of such devices are now accepted (De Graaf et al., 2015).

A body of empirical work has identified facilitators and barriers to technology acceptance. In the interview study (Mitzner et al., 2010) the main reasons older users liked technology were that they provided support for doing activities and convenience (as making life easier in some way), while the most frequent dislikes were inconvenience including perceived interruptions, financial expenses and effort, security and reliability. A similar study, but in the context of mobile devices (Müller et al., 2015), revealed that the acceptance process could be facilitated by a step-wise introduction of technology features; understanding the technology language against the background of older user practice; and social learning process among older users. The qualitative study described in (Waycott et al., 2016) examined older users' barriers to participation in a social isolation intervention. The intervention included an IPad application for creating and sharing photos within a closed social circle comprising a group of older users and their caregivers. The barriers were grouped as health conditions and difficulty learning new things (personal reasons), family circumstances and established social norms (social reasons), and language of use (technological reasons). Elers et al. (2018) conducted semi-structured interviews with older users and their informal support networks (friends, neighbors, and family members). The interviews concerned their needs for technological support to age in place. The participants expressed the needs for a low-cost technology; ease of use; personalized and controllable features (what information is collected and shared); and notification of the support network in case of emergencies or unusual activities such as the absence of movement in the home during certain hours.

In addition to this theoretical validation work, a stream of studies have investigated older users' attitudes and preferences towards different technologies, most recently robotics. This includes a cross-cultural interview study involving older users and caregivers to collect user requirements for a service robot that provides at-home assistance (Kleanthous et al., 2016). The study revealed that older users prefer a robot behaving as a friend but that they can control its behavior and one in which the benefits it brings are very apparent to them. Another research study examined older users' needs for the GiraffPlus system, an Intelligent Ambient Assisted Living (AAL) technology to support independence and improve quality of life (QoL) (Cesta et al., 2018). The GiraffPlus conducted unintrusive data collection (such as blood pressure, body temperature, movement, and fall), which are then analyzed to alert the caregivers to emergencies. The respondents highlighted the usefulness of physiological and home environment monitoring and facilitating social interactions. Amirabdollahian et al. (2013) investigated older users' perceptions of assistive robots in laboratory settings. They found that the acceptance of a robot is increased if the robot's personality (perceived through their appearance and behavior) matched users' expectations concerning a particular task.

A systematic review study identifies salient factors that influence older users' acceptance of technology (Lee and Coughlin, 2015). The factors cover individual, social, technological, and delivery aspects of technology acceptance and cover value, usability, affordability, accessibility, technical support, social support, emotion, independence, experience, and confidence.

Ensuring that older users are engaged early in the technology design to anticipate and reason about future roles that technology can acceptably play in their lives is still not commonplace. The previous theoretical and empirical work focused either on a specific technology (such as robotic technologies in Cesta et al., 2018, Heerink et al., 2010, Kleanthous et al., 2016; video games in De Schutter et al., 2015; and mobile social networking applications in Müller et al., 2015, Waycott et al., 2016) or ICT in general (Barnard et al., 2013; Elers et al., 2018; Lee and Coughlin, 2015; Macedo, 2017; Mitzner et al., 2010). In contrast, the ACANTO system is multifaceted, incorporating robotics, social networks, activities and recommender system that captures different well-being aspects at the same time, addressing both physical and mental health. This paper contributes a structured methodology to increase and improve older user involvement, which is necessary to achieve inclusive design that connects assistance and user experience, according to Farage et al. (2012).

*3.2. ICT for older users*

Research in computer science has overly focused on the negative effects of aging (Light et al., 2016; Vines et al., 2015a). Different application contexts have been targeted by a variety of technological systems, ranging from monitoring systems for fall prevention using wearable and ambient sensing technology (Danielsen et al., 2016), social robots for the well-being of people with dementia and mild cognitive impairments (Whelan et al., 2018) through to games for maintaining user engagement during therapy and rehabilitation (Uzor and Baillie, 2014). Despite this diversity, most of the systems share similar requirements. They were designed as assistive technology to compensate for the age-related decline in motoric, perceptual, and cognitive abilities, which make daily living activities more difficult for older adults (Cozza et al., 2017; Parra et al., 2014; Suwa et al., 2020; Yusif and Hafeez-Baig, 2016). A systematic review of ubiquitous technology designed for older users is provided by Cozza et al. (2017), in addition Gerling and colleagues (2017) describe 15 systems to facilitate physical exercise in older age which were published in the HCI literature.

Age-related declines in personal abilities can make daily living tasks more difficult for older adults (Mitzner et al., 2010). In this respect, robotic technologies have been widely used to assist older adults in different activities, mainly to support human physical actions, such as strength, balance, bending and endurance (Pu et al., 2019; Robinson et al., 2014). A review and meta-analysis of randomized controlled studies showed that these technologies are more effective in prevention than in rehabilitation (Pu et al., 2019). The review highlighted the contradictory nature of the evidence with some studies suggesting robots can reduce anxiety and loneliness and improve QoL for older adults. However, other studies indicated no significant influence on depression and quality of life. The review found no significant effect of social robots on cognition, potential for impact on quality of life, improved engagement and social interaction and had potential to reduce loneliness.

Robotic technologies can be designed to support people living independently by assisting with mobility, health monitoring, safety, or acting as companions. The former may be non-social robots, whereas the latter are social robots that have the main goal of offering companionship (e.g., PARO, the seal robot by McGlynn et al., 2017). Both types of robots (social and non-social) are nevertheless *assistive* robots and very little research has been conducted into designing robotic technology for enhancing the experiences of older adults. Different studies have investigated how to use robots as *companions* (Beer and Takayama, 2011; McGlynn et al., 2017; Orlandini et al., 2016) or *carebots* (Suwa et al., 2020) to assist older adults in various physical and social activities in home settings.

Remote telepresence robots can be useful for healthcare. The Giraff is a telepresence robot that uses a video interface to allow caregivers and relatives to virtually visit older people in their homes (Orlandini et al., 2016). The robot was evaluated in a 42-month longitudinal study, with different demographic groups (Italy, Spain, Sweden) including older users, family, and doctors. The participants highlighted the robot's usability and the appropriateness of video interface size, contents and interactions, and ergonomic features regarding camera position and orientation, adjustable size, and controllability (such as low light conditions and obstacle detection).

One of the challenges with designing healthcare robots is acceptance, as they are mainly designed to fix a problem rather than fully address user desires that might prevent decline and promote health (Robinson et al., 2014). On the other hand, robots may provide users with a way to communicate with friends or family enjoyably (Orlandini et al., 2016). This way, enhancing their communication experience outweighs the assistive nature of the technology (Vandemeulebroucke et al., 2018).

While it is essential to identify older users' problems and develop robotic technologies to solve these problems, much of our use of technology is in the realm of enhanced experience rather than assistance. Less attention has been given to the development of robots that support not only physical but social and emotional needs of older adults (Khosla et al., 2017). The acceptance of these technologies can depend on how they are presented to the user. If a robot is presented as assistive, it can evoke stigma (Wu et al., 2014) and thus discourage acceptance. People like to feel independent, and a robot that conveys their sense of need or disability is likely to be met with rejection or at least ambivalence. Matilda was a social robot with human attributes (such as baby-face-like appearance, human voices, gestures, and body movements) that could recognize voices and faces and perform activities such as playing music, dancing (as moving patterns), and playing card games. A 4-year study on acceptance of Matilda by older adults living with dementia (Khosla et al., 2017) revealed increased engagement with the robot concerning emotions (through reminiscence) and behavior (through pleasurable activities such as dance and games). Kachouie et al. (2014) conducted a systematic literature review on socially assistive robots for older adults. The majority of robots were animal-like, and the studies happened mainly in care facilities and nursing homes in Japan. They revealed that the robots capable of supporting broader aspects of older people's well-being were more acceptable than those with limited coverage.

Older adults appreciate supportive technologies generally but often cannot see a direct need for it in their own life. As elaborated by Cozza et al. (2017), the rejection is likely generated by a mismatch between the assumptions that have shaped technology design (focusing on functional aspects) and actual user needs/desires. While it is important to identify older users' problems and try to meet these needs, attention should also be paid to how technology can enhance life experiences rather than providing assistance *per se*. We need a more in-depth consideration of older adults' needs when designing the technologies to offer an experience-enhancing function rather than merely assistance (which connotes disability and dependence).

### *3.3. The model of user desires*

The model of desires for preferred activities (De Angeli et al., 2020) was derived from an interview study (N=18) performed at the beginning of the ACANTO research, whose methodology is described in Section 4 and Figure 3. The model was built on IBM that served as a theoretical foundation for creating the interview proforma. The interviews aimed to identify the activities preferred by older users and the factors that drove their intention to engage with those activities as desires. The model of user desires represents a grounding of IBM in the domain of the preferred activities in later life. The relations between IBM and the model of desires are illustrated in Figure 2.

IBM identifies the drivers of intention to engage in a given behavior in terms of the *attitudes*, *perceived norms* and *personal agency* towards that behavior. The *attitude* is a person's predisposition towards a particular behavior, determined by the emotional responses to undertaking the behavior (experiential), and the beliefs about the behavior's outcomes (instrumental). The *perceived norm* reflects the social influence on performing a behavior, based on what others think (injunctive norm) and what others are doing (descriptive norm). *Personal agency* consists of perceived control (the degree to which environmental factors may influence carrying out the behavior) and self-efficacy (the confidence in one's ability to perform the behavior).

According to the model of desires (Figure 2), preferred activities are influenced by the desire for *resilience* as the ability to recuperate fast from difficult situations (personal agency); the desire for *feeling good* and *doing good* (attitudes); the desire for *autonomy* and *belonging* (perceived norms).

*Resilience* requires *resources* and *competence*. *Resources* highlight a broader desire of older people to be of use to other people. When supporting their families or volunteering, they become a resource not only for themselves but for the community. *Competence* refers to the accumulated, life-course experience, knowledge and skills, a unique characteristic of later age for healthy people which older adults desire to pass on to others.

The attitudes are presented as desires for *feeling good* and *doing good*. They are the most influential factors of behavioral intention. *Feeling good* concerns pleasurable physical and mental experiences when engaging in preferred activities. *Doing good* is related to instrumental outcomes from performing activities which improve mental and physical well-being.

Perceived norms are presented as a balance between desires for decisional *autonomy* and *belonging* to social groups. *Autonomy* indicates the marginal influence of people who are not closely related to an older person. At the same time, preferred activities satisfy a desire for *belonging*, as family and friends exert normative influence on older people.

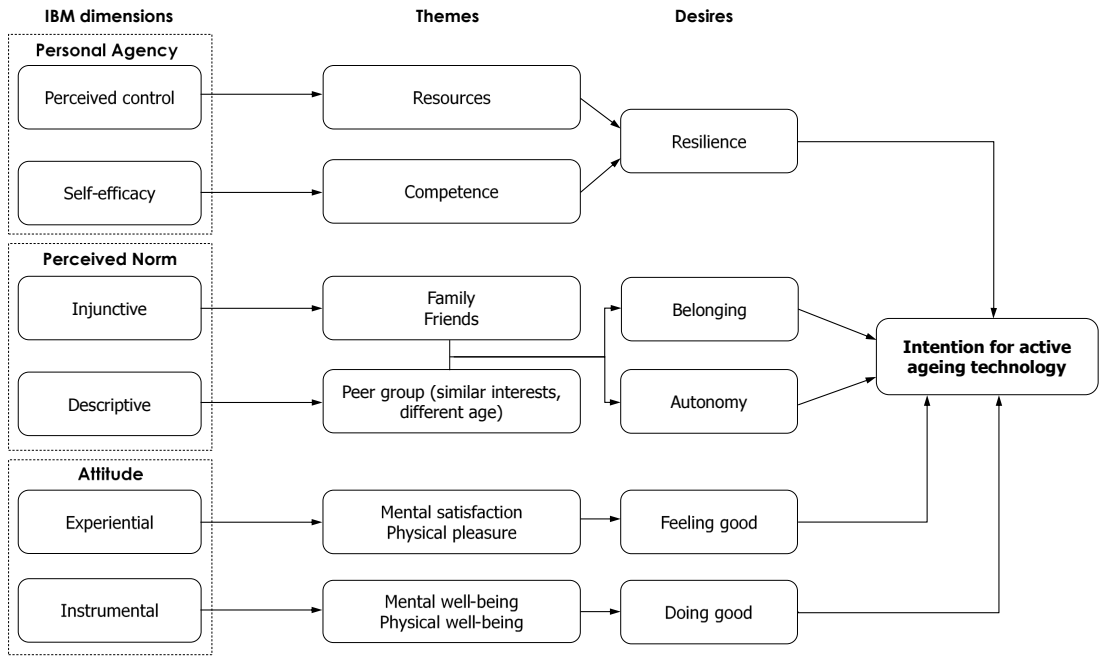

Figure 2. The model of desires for active aging technology (De Angeli et al., 2020). The IBM dimensions (left) were used when designing the interviews to elicit older users' preferred activities, and the factors that influenced the intention to participate in the activities as desires (right).

Based on the model of desires, we articulate design guidelines as *design for pleasure* and *design for resilience* (De Angeli et al., 2020). Design for pleasure refers to the need to include aesthetic features in assistive technology. Our previous study (De Angeli et al., 2020) revealed that pleasure and enjoyment are important motivators for older people. Elements of aesthetics, sensory experience and feelings define prospective design spaces for technology in later life. Design for resilience highlights older people as resources for the community and their competence developed through their lived experience. That way, the focus of the design moves from the implementation of assistive technology to the creation of tools that promote pleasure, self-efficacy and increase the quality of later life (Light et al., 2015; Light et al., 2016; Vines et al., 2015a; Yuan et al., 2018). It presents an affirmative picture of older adults as resourceful and resilient, emphasising their knowledge and skills while respecting their desires.

We use this person-centric model of desires to ensure older users are engaged from the early phase of the design in understanding and reasoning about the future system.

**4. Methodology**

The research was organized into four steps, as illustrated in Figure 3. Phase 1 used the IBM model as a framework to identify the activities preferred by older adults and the elements that affect their intention to engage with those activities as desires (De Angeli et al., 2020). This phase resulted in the model of user desires (described in Figure 2 from section 3.3). Phase 2 created a set of user scenarios in which the ACANTO system was presented as supporting the different preferred activities. Phase 3 evaluated the scenarios concerning the model of user desires. The model served as a tool to operationalize interviews to elicit user requirements (the questions are reported in the Appendix). Phase 4 developed the study's findings into a derived model that describes user requirements for active aging technology. This paper describes the work from Phases 2-4, from which we derive the model of user requirements.

The research was approved by the Ethics Committee for the experimentation with human beings at the University of Trento (Italy) and the University of Northumbria in Newcastle (UK).

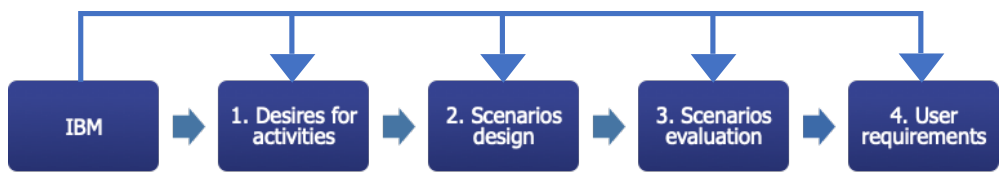

Figure 3. Structured research methodology for user requirements elicitation.

The principal way in which the ACANTO system facilitates well-being is to suggest activities and support the user to take part in them. Collecting information about interests, background and previously enjoyed activities from the users was crucial to the formulation of recommendations. Frequently reported activities from Phase 1, previously reported in De Angeli et al. (2020) included physical activity (e.g., walking), volunteering, hobbies (e.g., shopping, lifelong learning, museums) and socializing (e.g., visiting friends or family). The next step was to investigate the role of ACANTO system should play in planning and/or performing future activities. We adopted a scenario-based requirements approach for data collection (Sutcliffe, 2003) to contextualize different aspects of living with and using the ACANTO system, combining physical and social elements. We designed a set of 14 scenarios to demonstrate different ways the system can be used to support the particular activities extracted from the previous study. A scenario is designed as a hypothesized story describing typical users undertaking their preferred activities (Preece et al., 2015) with the help of the ACANTO system. The system assumes two distinct contexts of use – for older adults living independently, without medical supervision, to increase their daily activities within a living environment, and in a hospital or care facility under medical supervision. Accordingly, we have two groups of scenarios – well-being and rehabilitation. Each scenario is centered around a persona and an activity. We considered natural changes in health and functional capacities that may appear in later life in creating the set of personas. We included social, economic, and environmental factors that may affect participation in the activities. The resulting set contained personas with different socio-economic statuses and a range of functional capacities, including healthy older adults and others who experienced a decline and wished to regain the desired health and mobility level. Table 1 summarizes each persona's demographics, health, and lifestyle.

The scenarios use desires and related themes to represent the factors that may influence the intention to engage in the activities. These scenarios present a range of hedonistic activities (e.g., socializing and learning) from which users may derive pleasure and satisfaction through to practical activities (e.g., exercises) from which user derive instrumental benefits (e.g., mobility and autonomy). Table 2 describes the scenarios in terms of personas, activities, user desires, and the ACANTO system components used (FriWalk and FriTab). In addition to the model of user desires, we introduced additional categories within the perceived norms construct to examine the influence they may have on user intention as professionals (injunctive influence) and peers (descriptive influence).

The scenario structure is organized into four main parts:

- information about the existing activity (lifestyle and gap for activity);
- information on how the activity might be enhanced (desire for activity);
- an example of how the ACANTO system can support the enhancement (arrival of the system, the activity recommendation, and support for the activity execution); and
- the user's response to the recommendation and the expected outcome of using the ACANTO system (result).

Each scenario starts with a brief description of the persona, including demographic information, location, and lifestyle (Table 1). It continues with a description of the persona's need for the activity. Then the scenario shows how the persona acquires the ACANTO system. For example, by approaching the persona directly or from the people (s)he knows. The persona starts using the FriTab to provide personal data (e.g., location, background, or preferences for activities) or to search for activities or people of interest. Based on the input, the persona is informed about an opportunity for the activity. Following the FriTab's recommendation, the persona decides whether to accept it. Depending on the persona and the activity, the recommendation can be implemented with or without the FriWalk (i.e., robotic walker) support. The scenario concludes with doing the activity and a description of the benefits derived.

Each scenario involves a different persona (Table 1). Tom (Scenario 1) volunteers in a museum, tries out the FriTab that suggests people interested in natural history, and organizes a guided tour with the ACANTO system's help. George (2) uses the FriTab, which offers him an acting group to join and work with various-age members. Anthony (3) is introduced to the FriTab that suggests new people to meet according to their shared background based on the data he entered. Isabella (4) is presented with the ACANTO system at the shopping mall; the system recommends and guides her to her preferred shops. Sarah (5) receives the FriTab that suggests attending the University of the Third Age (U3A) together with a friend (she accepts and visits). Michael (6) was given the ACANTO system that recommends and assists in a guided museum tour that he attends. Isabel (7) receives the ACANTO system that suggests and supports meeting a new person with shared interests (walking) with whom she spends time. Fatima (8) uses the ACANTO system in moving around; the system detects a mobility issue and recommends a doctor visit that she carries out. Dorothy (9) is introduced to the ACANTO system that suggests her going out and helps her moving around. Scenarios 10-14 describe personas with different life circumstances who wish to regain mobility after experiencing a functional decline (due to impaired eyesight or fall) - the ACANTO system assists in walking through rehabilitation or moving outside. Table 2 describes the scenarios concerning user desires, indicating personas, activities, and ACANTO system components (FriWalk and FriTab).

For our study, we implemented the scenarios as animated storyboards by following the guidelines in terms of the level of detail, the amount of text, the presence of people and emotions, and the length, as suggested in (Truong et al., 2006). The mean duration of the scenarios is 93 seconds (Max 115 seconds, Min 61 seconds). Figure 4 illustrates the storyboard excerpts from the scenarios, whereas thorough video descriptions of the recorded scenarios as animated storyboards can be found elsewhere [2].

---

[2] Recorded video descriptions of the scenarios as animated storyboards: http://bit.ly/2sCGEVd

Table 1. Personas described in terms of demographics, location and lifestyle. Taken from McNeill et al. (2017), p. 12.

| No. | Description of persona |
|---|---|
| 1 | *Tom* is a 68-year-old man who lives with his wife. Tom lives in Byker, Newcastle. Tom has quite an active social life and enjoys meeting new people. |
| 2 | *George* is an 81-year-old man who lives alone. He lives in Newcastle. George is quite active and enjoys helping others. |
| 3 | *Anthony* is a 75-year-old man who lives alone. He lives in sheltered housing near Gateshead town center. He lives a quiet life and doesn't get out much. Two or three times a week he walks to the local superstore to buy groceries. |
| 4 | *Isabella* is an 89-year-old woman who lives alone. She lives in Newcastle City Center. She is not as active as she used to be because her eyesight is getting worse and she worries about losing her way. |
| 5 | *Sarah* is a 72-year-old woman who lives with her husband. She lives in a small house in Corbridge. Sarah enjoys living in the rural community and has a small number of friends who live nearby. |
| 6 | *Michael* is a 72-year-old man who lives alone. Michael lives in Felling, Gateshead. For the past few years, he has found mobility very difficult and he is waiting for a hip operation. Consequently, he doesn't get out much. |
| 7 | *Isabel* is an 82-year-old woman who has lived alone for the past two years. She lives in a flat by herself in Newcastle. She no longer goes out very often and has become very physically inactive. Her daughter brings her groceries once a week. |
| 8 | *Fatima* is a 73-year-old lady who lives with her husband. She lives in Hexham. She is reasonably active and enjoys using her walker to help her get around. |
| 9 | *Dorothy* is a 69-year-old woman who lives alone. She lives in Blaydon. Dorothy uses a walker to get around because she finds that it gives her confidence after her fall one year ago. |
| 10 | *Manuel* is a 74-year-old man who lives alone. He lives in an apartment in Madrid. He recently suffered a bad fall which has made it difficult for him to get around. He is undergoing therapy at a local falls clinic and is slowly getting better. |
| 11 | *Manuela* is an 83-year-old woman who lives with her husband in a nursing home in Getafe, Madrid. She has lost some sight during the last years, and she does not know her way around the new neighbourhood or the venues and activities there. |
| 12 | *Jose* is an 80-year-old man who lives alone. He lives in a small, old flat in Getafe, Madrid. José was quite active before the fracture, and enjoyed going in the morning to a bar near his house with his friends. |
| 13 | *David* is a 75-year-old man who has recovered from a fracture and currently he is able to walk assisted by a normal walker. |
| 14 | *Ana* is a 76-year-old woman who has some blood pressure problems. She lives in a nursing home and needs a crutch to walk after a fall. |

Table 2. Descriptions of the scenarios concerning personas, activities, desires, and related themes to introduce the factors that may influence the intention to engage in the activities and ACANTO system features. +/- denotes the presence/absence of the specific theme in a scenario. When used in the Belonging theme (age columns), +/- indicates the presence/absence of age groups different than that of a persona (as ≠) or age peers (as =). Time discussed is the number of times a scenario was chosen and discussed by the participants.

| No. | Persona | Activity | Attitude | | | | Norm as a balance between belonging and autonomy | | | | | Personal agency | | System feature used | Times discussed |
|---|---|---|---|---|---|---|---|---|---|---|---|---|---|---|---|
| | | | Feeling good | | Doing good | | Autonomy | | Belonging | | | Resilience | | | |
| | | | mental satisfaction | physical pleasure | mental well-being | physical well-being | friends | professionals | interest | age = | age ≠ | Resources | Competence | | |
| 1 | Tom | volunteering in a museum | + | — | + | — | — | — | + | + | + | + | + | FriTab | 10 |
| 2 | George | volunteering in an acting group | + | — | + | — | — | — | + | + | + | + | + | FriTab | 9 |
| 3 | Anthony | spending time with others | + | — | + | — | — | — | + | + | — | + | + | FriTab | 7 |
| 4 | Isabella | exploring shopping mall | + | — | + | — | — | + | — | — | — | — | — | FriTab, FriWalk | 6 |
| 5 | Sarah | attending U3A | + | — | + | — | + | + | + | + | — | — | — | FriTab | 4 |
| 6 | Michael | visiting museum | + | — | + | — | — | + | + | + | + | — | — | FriTab, FriWalk | 4 |
| 7 | Isabel | walking | + | + | + | + | — | — | + | + | — | — | — | FriTab, FriWalk | 3 |
| 8 | Fatima | shopping | + | — | + | — | — | + | — | — | — | + | + | FriTab, FriWalk | 3 |
| 9 | Dorothy | shopping | + | — | + | — | + | + | — | + | — | + | — | FriTab, FriWalk | 3 |
| 10 | Manuel | visiting museum | + | — | + | — | + | — | — | — | — | — | — | FriTab, FriWalk | 3 |
| 11 | Manuela | rehabilitation exercise | + | + | + | + | — | + | + | + | — | — | — | FriTab, FriWalk | 2 |
| 12 | Jose | rehabilitation exercise | — | + | — | + | — | + | — | — | — | — | — | FriTab, FriWalk | 2 |
| 13 | David | visiting museum | + | + | + | + | — | — | — | — | — | — | — | FriTab, FriWalk | 1 |
| 14 | Ana | rehabilitation exercise | — | + | — | + | — | + | — | — | — | — | — | FriTab, FriWalk | 0 |

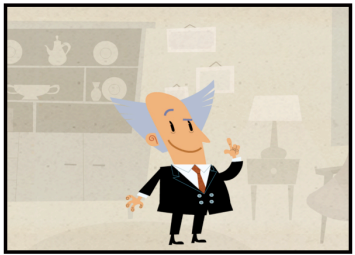 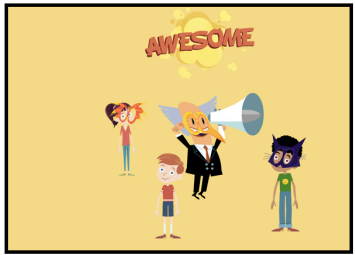 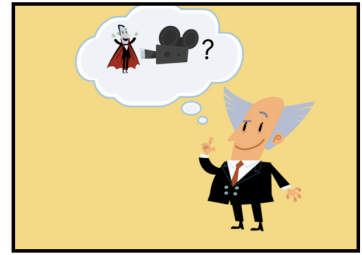

George is a 81-year-old man who lives alone, he lives in Newcastle.

George is quite active and enjoys helping others. He has a keen interest in acting and used to be a member of an acting group. He enjoys helping younger people to develop their skill.

One day George is introduced to the FriTab system and starts to use it regularly, he uses it to find out about local events. He would like to be a member of an acting group but is not aware of any nearby.

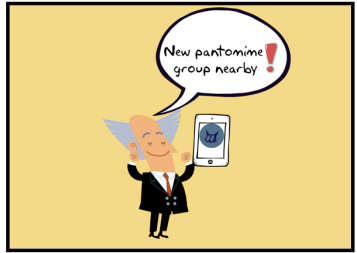 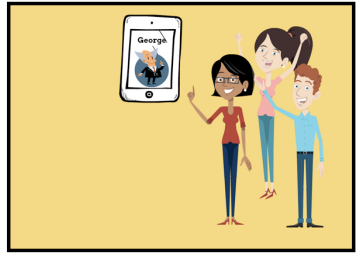 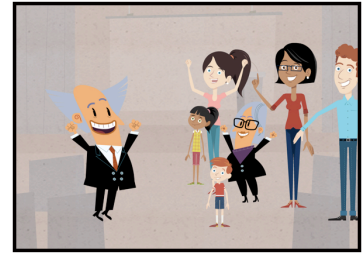

One day when George is checking his Fritab, he discovers that there is a new pantomime group starting at a nearby small theatre. The system suggests that he join the group based on his interest in acting. George accepts the sytem's recommendation and it sends a message to the acting group confirming his attendance to a meetup the group is planning.

The Fritab alerts the rest of the group that George will attend. They tell George that they will meet in the following week on Friday at 1:00 PM in the local small theatre.

George attends the meetup and finds that the acting group is made up of diverse interesting people, including younger people whom he enjoys sharing his skills with.

(a)

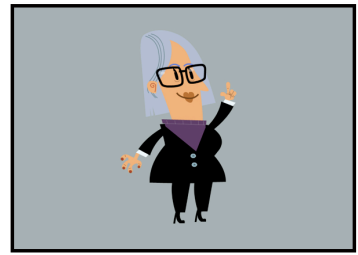 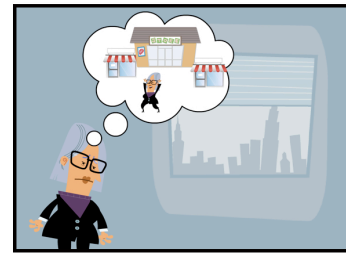 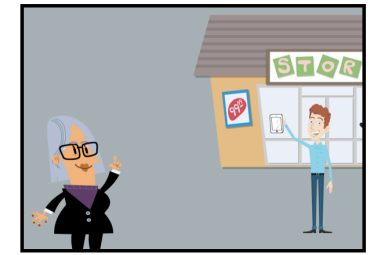

Isabella is a 89-year-old woman who lives alone in Newcastle City Centre. She is not as active as she used to be because her eyesight is getting worse and she worries about losing her way.

Isabella likes to shop and would like to visit large shopping centres without being afraid of getting lost.

One day she visits a shopping center and a member of the staff suggests that she try the new FriTab and FriWalk system. She doesn't have her own Fritab so the shopping center provides one.

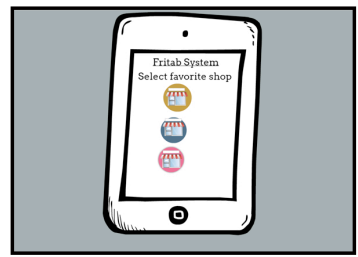 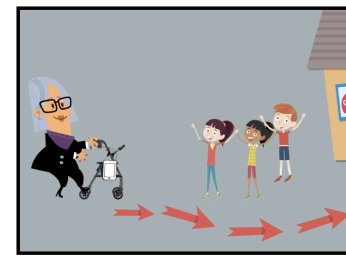 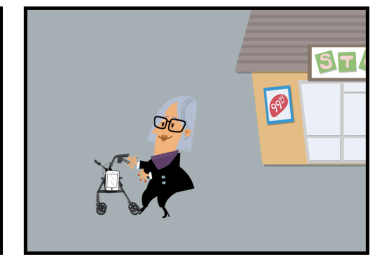

She selects her favorite shops from a list and the system shows her directions of how to get there. After visiting her favorite shops, the system recommends other shops she might be interested in. The system guides her to the new shops.

While approaching the shops the system sees that there is a group of young people ahead of her and it gently steers her around them.

After exploring the shopping centre with the help of the FriWalk, she is impressed with the abilities of the system and thinks she will use it again.

(b)

Figure 4. Storyboards' excerpts of the scenarios: (a) Scenario 2 – George using FriTab to meet new people; (b) Scenario 4 - Isabella using FriTab and FriWalk to navigate in a shopping center.

## 5. System scenarios evaluation

The model of desires guided the design of an interview study to examine the factors that affected older adults' intention to use the ACANTO system to undertake their preferred activities, as described in the scenarios.

The study was organized as semi-structured interviews with 21 participants. They were from Newcastle, UK (11 females, 10 males; age range 60-89, mean = 74, standard deviation = 8). The sample was recruited from the NorthEast Age Database at Northumbria University. It contained people from an urban area, living independently at home, and did not suffer any severe health issues. Around 14% of the Newcastle's population were over 65 years, expected to increase to 44% by 2039 (Well-being for life, 2017). The aging index (the number of older adults per 100 persons younger than 15 years old as from Preedy, 2010) based on 2011 census data is approximately 81% (Nomis, 2017). The participants had experience with smartphones or tablets for emails, messaging, and online applications (such as public services, booking, social networks, and media content). Our previous study (De Angeli et al., 2020) engaged N=18 independently-living users from the UK (8) and Italy (10). The participants recruited for this study were different persons than those recruited in the previous one, but had similar technology experience and health (as above).

The interview was pilot tested with 1 participant. The pilot interview led to the refinement of the initial questions and the coding scheme (derived from the model of user desires described in Table 3). The final interview questions are reported in the Appendix. The interviews lasted 50 minutes on average. A male researcher ran all the interviews. The participants were interviewed at the PaCT Lab, Northumbria University.

*5.1. Procedure*

At the beginning of each interview, participants were briefed about the ACANTO system, its main components, and their purpose. At this stage, we described the FriWalk and FriTab components, their functions, and how they could be used without going into detail about the technology or scenarios. When describing the components (and creating narratives of the scenarios' animated storyboards), we did not use technical language but terms that the participants could understand.

The study's aim was described to the participants as collecting information on people's intentions to use the system, participants were provided with information about the procedure, including recording for analysis and how their anonymity would be maintained. They were then invited to sign a consent form.

The interview started with the questions regarding the participant's lifestyle to understand the personal context. The participants were presented with cards that describe the personas from Table 1. They were asked to choose up to 4 personas they most identified with or that were similar to someone they knew well. By choosing personas similar to themselves or other people they knew, the participants were best placed to determine the personas' intentions to use the ACANTO system or their anticipated experiences of using it. The majority of the participants chose 2 (N=8) or 3 personas (N=8), 4 participants chose 4 personas and 1 participant chose 1 persona.

For each persona, participants watched the associated animated storyboard of the scenario once. All storyboards are available from http://bit.ly/2sCGEVd. Then the animated storyboard was repeated and paused at two specific points. Firstly, after introducing the persona, the researcher asked the participant about the facilitators and barriers associated with the persona that influence the use of ACANTO system (i.e., FriTab and/or FriWalk used in the particular scenario). Secondly, after demonstrating the ACANTO system's role, they were asked to elaborate on the persona reasons for use of the system. These questions concerned what is it about the scenario that would make the persona feel good (experiential attitude), what might the persona gain (instrumental attitude) or be able to do (doing good), how would other peoples views about or use of the system affect their behavior, did the system promote autonomy and belonging (normative influences), build resources and facilitate competence (agency).

*5.2. Analysis*

We audio-recorded and transcribed the interviews. The participants were anonymized by a string of letters and numbers (participant number and sex) and number (age). The transcripts were processed using deductive thematic analysis (Braun and Clarke, 2006). The analysis was conducted by identifying and discussing all the statements relevant to the coding scheme. It was supported by the Atlas.ti software. Four code families were derived from the model of user desires (Figure 2). In particular, three corresponded to the original dimensions of the IBM (attitudes, norms, and agency) and their respective themes as from the model, one related to barriers. Each code family (as from IBM) was further classified into desire-specific codes, as reported in Table 3. Two independent analyses were carried out, one in Italy one in the UK in parallel to ensure agreement with the coding. They confirmed that the data was complete, and there were no disagreements (with the inter-rater reliability of 85%). The analyses of the transcripts against the codes discovered the themes presented in Table 3.

*5.3. Results*

The last column in Table 2 shows how many times each persona (scenario) was chosen and discussed with the participants. As participants could choose more than once, 57 scenario discussions were completed in total. Participants tended to identify with the more active persona scenarios. Tom was the most popular - showing an engaging person who likes meeting new people and helping others. 10 out of 21 participants chose this persona scenario which involved personal involvement in diverse volunteering activities. They perceived the system to enable them to meet new people to socialize with or help. The most discussed scenario using the FriWalk is Isabella, a generally healthy woman but experiences sight issues. Visual capability typically decreases as people age. 6 out of 21 participants who identified with Isabella also faced an eyesight issue. They considered the system would support normal functioning with this condition. The main problems due to the eyesight included loss of confidence in going out, difficulty using public transportation, and finding items in shops. The participants who selected rehabilitation scenarios did so from the perspective of a friend or family member. These participants typically had mobility problems or were not active themselves.

Table 3. Themes from the evaluation of the ACANTO system scenarios, with the number of times the theme occurred in participants' responses and summative values for codes and code families.

| Code family | Codes | Theme | Count | Total | |
|---|---|---|---|---|---|
| Attitudes | Experiential (Feeling good) | *mental satisfaction* | 29 | 47 | **121** |
| | | *physical pleasure* | 18 | | |
| | Instrumental (Doing good) | *socializing* | 48 | 74 | |
| | | *volunteering* | 13 | | |
| | | *utility* | 13 | | |
| Norms | Injunctive (Significant others) | *family* | 18 | 39 | **77** |
| | | *friends* | 12 | | |
| | | *professionals* | 9 | | |
| | Descriptive (Reference group) | *similar others (mixed age)* | 38 | 38 | |
| Agency | Perceived control (Resources) | *ergonomics* | 12 | 36 | **144** |
| | | *reliability* | 10 | | |
| | | *function quality* | 8 | | |
| | | *availability* | 6 | | |
| | Self-efficacy (Competence) | *usability* | 77 | 108 | |
| | | *mobility* | 20 | | |
| | | *benefits* | 11 | | |
| Barriers | | *health* | 20 | **35** | |
| | | *cost* | 10 | | |
| | | *learning effort* | 5 | | |

In this section, we now present our findings according to the coding scheme, describe themes that emerged from the data, and illustrate them with excerpts from the interviews. Each quote is identified with the letter initials of the participant, sex, age, and name of the persona/scenario. The number of occurrences of the particular theme in participants' responses is given in parentheses, and the same values are reported in Table 3 (*Count* column).

*5.3.1. Agency*

Ergonomic design (12 occurrences) showed a major influence on the users' self-perception of being resourceful. It was elaborated in terms of adjustable size (to prevent slouching), grips (handbrake sensitivity), transportability, ease of manoeuvre (slim) and using non-embarrassing signals.

Other relevant factors include reliability (e.g., braking system) (10), the function quality (8) described as the accuracy of activity recommendation in providing relevant information (e.g., cost, weather, accessibility, and transportation options) and appropriate matching when recommending, and availability of technology to older users (6).

The participants' competence to use the system (self-efficacy) was discussed concerning the three main themes - usability, mobility, and benefits.

Firstly, it concerned usability. Participants frequently reported that the simpler and easier to use the system is, the more likely people are to feel competent using it (77 occurrences):

> I would try everything you wanted me to [on the system] out of curiosity but to keep doing after that it would have to be fairly easy to use. If it wasn't [fairly easy to use] then I would lose interest in it because I would be doing something that until I knew how it worked and what it did, I wasn't interested in. [P1M70-Sarah]

Instructions and explanations were the main factors of contributing to a sense of competence (20). These were mentioned as initial assistance/education about the system, step-by-step instructions, peer help, manuals and handbooks. The technology's language should be understandable, using clearly-worded instructions (10). When asked about being in control of the system, a participant explained:

> Your technical language would make me struggle. I wouldn't understand it. I assess lay summaries sometimes and I look at what they have written and I say to them, "Is this a lay summary for someone like me or is it a technical summary for your professor and that?" [P2F72-George]

Secondly, the system was perceived as a means to facilitate competence concerning one's mobility (e.g., going out and getting around) (20):

> I think that [using the Friwalk] would be really handy for me, just a little - but not drag me like, because I have floppy ankles, so I have to make sure I pick my feet up. [P3F78-Dorothy]

Thirdly, the benefits of using the system need to be evident for older adults to increase their sense of competence (11). It may be challenging to get older people to use technology for the first time. Still, the participants highlighted that if the advantages of using it are apparent, they will be more inclined to invest in learning and overcoming obstacles to usage. While describing his confidence in using the system, a participant commented:

> I think it [running a computer] depends on the way that it's done, but if you can show people the benefit then they're generally willing to invest the work involved in understanding the system. [P4F71-George]

Nevertheless, positive comments towards the agency assume that a user should not depend too much on the system. For example, feeling a loss of autonomy or being "nannied" as if the system steered them around. Participants elaborated on specific needs for such a system concerning aging (e.g., fear of falling, lack of motivation, and confidence to be physically active). Relatedly, participants elaborated on specific resources such as having a seat on which they could rest and a basket to carry items (e.g., when shopping). This way, they may walk where they want without worrying about getting tired.

While some participants expressed this fear of stigma from the FriWalk, they also felt positive about it. For example, being an alternative better than a wheelchair since the user would be at the same height as other people, thereby interacting better. Or re-framing it as an enabling device that raises their competence:

> I know that some people might initially feel a bit embarrassed that they look like they had a tripod walker but if people all sort of look and they say, "This is my satnav," then you think, "Oh it's street cred, satnav, how does that work?" This is what you do with it and it makes you feel worthy because you're actually telling somebody who's probably a lot younger than you how this works. So that's a street cred thing. [P5F69-Isabel]

Regarding *barriers* to the intention to use the system that affected agency, the participants elaborated on health and mobility issues (20) that included limited eyesight, walking uphill/downhill, hip and knee problems, and fear of falling. Furthermore, they highlighted they expected the system to be costly (10) and they had difficulty learning new things (5).

*5.3.2. Attitudes*

Findings on experiential attitudes (*feeling good*) show consistent entanglement of physical and mental experiences from the participants' narratives. Participants highlighted that the system would provide mental satisfaction (29 occurrences). In particular, they associated altruistic scenarios with feelings of enjoyment and achievement from helping others and recruiting new people to volunteer. The specific feelings reported by the participants were a sense of social involvement and belonging to a peer group (16), a sense of fulfilling life (5), a sense of achievement (4), and a feeling of independence (4). Physical pleasure was described as the feeling of relaxation (10) and physical well-being (8), such as feeling good or staying fit. For example, a participant raised that if the system is easy enough to learn, this will provide a feeling of "mastery", a psychological need connected to subjective well-being (Tay and Diener, 2011):

> I think for a start once they'd mastered the fact that they could actually use the FriTab or tablet or whatever, that would make them chuffed to bits, because I know how I feel when I can't [do a] slight thing now, you know. But again knowing that they could [learn it], it improves you both socially, mentally and physically when you get out and about from your usual boring day-to-day routine. [P6F69-Anthony]

In addition, expanding social circles by meeting people with similar interests or backgrounds through shared activities was seen as the primary *doing good* for themselves that the system may enable (33). Frequently mentioned activities were walking, social drink and food, a theatre visit, and shopping. The participants spoke about feeling connected by maintaining contacts with existing friends and family, creating new relationships depending on location and interests, and connecting with people they used to work with.

> It [system] would make them feel as though they weren't alone in the world. I mean he doesn't sort of say it in those terms but you can feel where's the world gone, you know, we're stuck out here. Obviously they know their next door neighbour either side but that's about it because nobody ever goes out anywhere. [P7M81-Sarah]

Helping other people emerged as an altruistic opportunity that the system may provide as *doing good* for others (13). The participants associated the volunteering scenarios with different kinds of such activities from their lives. These included charity work, teaching a language group, helping people with disabilities with their legal rights, organising group sessions for people with eyesight problems, local community work, tutoring at the U3A, and planning get-togethers. A participant elaborated on engaging younger people in volunteering activities:

> [...] so that [the system] would be great to fill the days [...] I volunteer for Age UK on the IT side [...] And on a Wednesday afternoon I go to one of the schools and do it there with sixth form pupils as a mentor. And they in turn work with the senior citizens, so it's the two-way street. [P1M70-Tom]

Connecting to people with similar health conditions as a benefit emerged in the rehabilitation scenarios (15).

Finally, participants highlighted the system's utility concerning *doing good*. In particular, helping them to get desired information related to the activities, including shopping (sales), transportation arrangements (bus stops and times), public services, and navigation in public spaces, including museums and shopping centers (13). A participant reflected on finding co-located shops and places to eat:

> [...] Could be nice not just shopping, but somewhere to eat. You know, I think that might be something you wouldn't know. So if it could direct you to the kind of food you want, that would be even better than the shops. [P8M78-Dorothy]

A barrier concerning attitudes was the feeling of being "past it". A participant mentioned his friends might say that they do not want to use technology, or to get more socially involved and may need more encouragement to go out.

*5.3.3. Norms*

Regarding significant others, the participants reported that they would likely use the system if encouraged by family members (18 occurrences) including a spouse, children, other close relatives and friends (12) whom they described as people they knew well and whom they trusted.

In addition participants highlighted a role for two different types of professionals. Firstly they suggested they would follow a recommendation provided by healthcare staff (4). Finally, their intention to use was influenced by the availability of a technology expert to adequately informed them about how to use it (5).

These last two groups illustrate the different types of people who may encourage an older adult to use system. In addition, participants also highlighted the influence of the descriptive (reference group). This includes people with similar age, health or background, or similar interests regardless of age. In this respect, the participants mentioned that they would use the system if they saw people similar to themselves using it (38). The system was recognized as a tool that can link people with similar health conditions (rehabilitation scenarios) or shared interests and background (well-being scenarios) together, that people may aspire to use to be part of the group and feel they were missing out if they too did not have one. However, the participants would only be encouraged to use the system as if other people communicated positive experiences of using it:

> It's like all of these. Things like Facebook. If you think, "Why is Facebook so popular?" It's because people started using it, and found that they couldn't communicate with their group of friends unless they were on this thing. There was a sense of missing out. So I think you need that sort of impetus into the start of the system. [P9M71-George]

In addition to people similar to themselves, participants also mentioned a desire to engage in intergenerational contact around volunteering scenarios, as illustrated in the quote [P1M70-Tom] about volunteering. The George persona, an older man helping young people in an acting group, was the second most discussed persona. However, intergenerational contact depends on the type of activity:

> I would have thought that possibly an age limit may be useful, but not necessarily because I guess with a thing like that scenario I guess it's the subject [acting] rather than the people [...] for this subject it probably wouldn't matter but there may be other things where you possibly wouldn't want to be meeting with a 20-year-old versus an older person. For various reasons. [P10F84-Tom]

## 6. Model of user requirements

We aggregated recurring themes from the data and collated them into the model of user requirements for active aging technology (Figure 5). The instrument for our analysis was the IBM-based model of older user desires (De Angeli et al., 2020). This model builds on the generic factors provided in IBM by providing concrete factors which specifically motivate older people to engage with technology to maintain an active lifestyle. It makes a shift from a perspective that connects later life with decline and frailty. Rather, it supports older users as people with not only physical needs to accommodate but with desires for pleasure and to give back to the community. This requires an inclusive approach to improve later life quality. As such agency was the dominant construct discussed as influencing behavioral intention (144 occurrences) followed by attitudes (121) and norms (77), as summarized in Table 3.

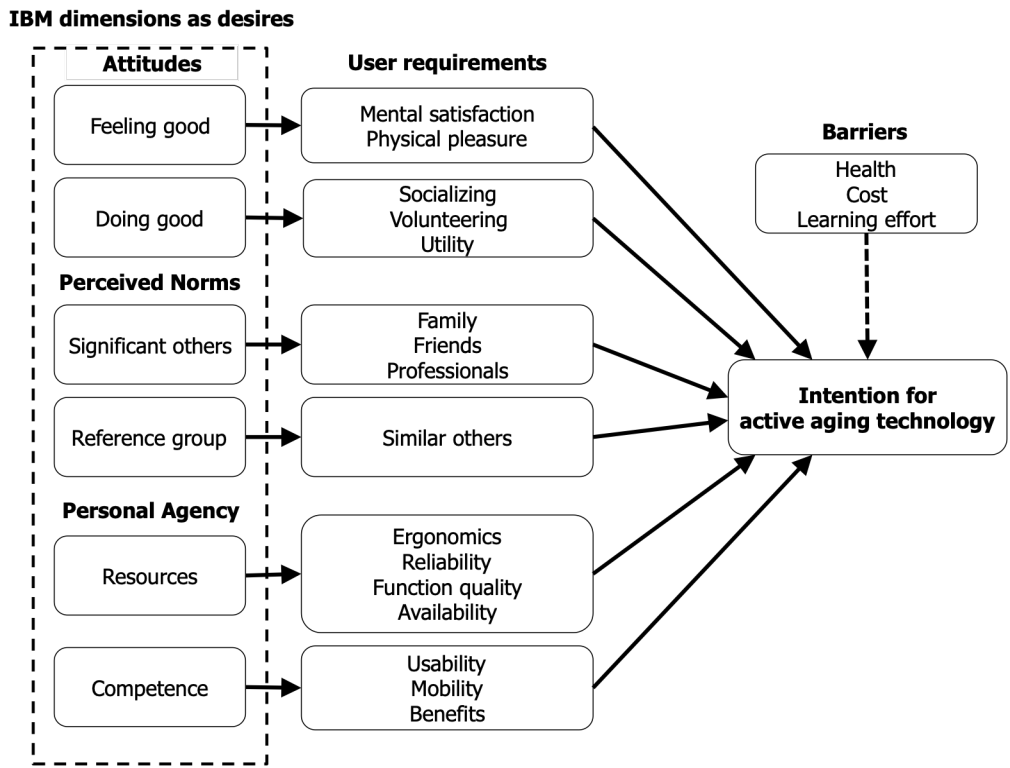

Figure 5. User requirements for active aging technology through the lens of the model of desires.

## 6.1. Agency

*Personal agency* consists of perceived control over the environment or tool and self-efficacy or confidence in ones ability. This requires resources and competence which participants relate to the technological properties. In particular, resources as a function of *ergonomics*, *reliability*, *function quality* and *availability*.

Participants highlighted some specific ergonomic design elements for robotic walkers (such as size, grips, controllability) that must consider natural physical and mental functioning changes as people age. Reliability as the system's quality was concerned with the walker-related aspects, while function quality was mainly referred to application aspects (i.e., recommender system).

Competence was facilitated by expectations of *usability*, *mobility* and *benefits* from using technology. Usability linked to participants' competence in using the system.

It was the most frequently referred to as ease of use, older-friendly instructions and support for learning, and clearly-worded language of use. The technology that facilitates mobility such as leaving home and spending time raises older adults' perception of their competency. Finally, the benefits of using technology in relation to achieving their personal goals must be apparent for older users.

*6.2. Attitudes*

Older adults' intentions are justified for the sake of *feeling good* and *doing good*. Findings on doing good indicate the entanglement of *physical pleasure* and *mental satisfaction* throughout the scenarios. Mental satisfaction came out from feelings of enjoyment, including connectedness, fulfillment, achievement and independence. At the same time, the physical pleasure was described as feeling good and relaxed.

The *doing good* resulting from technology use were *socializing*, characterized by interests, health and age groups; *volunteering* - helping others as a service/opportunity that technology enables, and *utility* as the system's usefulness in supporting preferred activities.

An important contribution of this study is identifying *volunteering* as a prominent activity among older people. It is perceived as an opportunity to provide support to older and younger people, and to the community in general. Volunteering allows older adults to use their knowledge and skills to demonstrate competence. The volunteering behavior of older adults is connected with higher levels of well-being (Anderson et al., 2009), positive health outcomes (Kim and Konrath, 2016; Varma et al., 2016), and decreased use of health and care services (Kim and Konrath, 2016).

*6.3. Norms*

Our findings are consistent with the previous work, where descriptive norms were perceived as more influential than injunctive norms (De Angeli et al., 2020). The structure of social influence is described with the specific networks of users. *Family* and *friends* had the major injunctive influence (as significant others), while the influence of *professionals* quoted as healthcare staff and technology experts was very moderate.

Descriptive norms were characterized as positive reference group experiences defined as *similar others* concerning interests and hobbies (involving different age groups), background, and health. Linking up with people recovering from similar health issues for support and encouragement was common to rehabilitation scenarios.

Lastly, the main barriers that interfere with older adults' intentions are *health issues*, technology *costs*, and difficulties in *learning new things*.

## 7. Discussion

The presented model is descriptive, proposing dimensions (as model's factors) to be taken into account when investigating user requirements rather than focusing on user requirements *per se*. Considering that user requirements are specific to a particular technological system, we suggest general dimensions in which concrete user requirements can be instantiated and categorized instead of collecting them *ad hoc*. The relevance of the dimensions can vary based on the technology for the older user. That said, the purpose of our work is a comprehensive model for eliciting user requirements. In this sense, user requirements are a means for validating the proposed model.

### *7.1. User requirements implications*

When people reach a certain age, psychologically and physically, they are content to keep things going rather than changing. Overcoming this remains a challenge. Granted that there are difficulties for older adults using technology by virtue of their age, a solution may be to target people before they reach that stage or design technology as age-friendly as possible. Our results add to existing critics of design for active aging (Vines et al., 2015a) and show that older adults' motivations for using technology cannot be easily discriminated from that of younger people.

*7.1.1. Implications for user requirements of agency*

The technology acceptance models' usability (TAM's perceived ease of use and UTAUT's effort expectancy), as a critical factor of technology use (Barnard et al., 2013), relates to the agency, enhancing older users' competence and themselves as resources for society. The usability is a common theme in empirical studies often referred to as old age-friendly language of use (Müller et al., 2015; Waycott et al., 2016) and optimized learning effort (Elers et al., 2018; Lee and Coughlin, 2015; Mitzner et al., 2010; Vassli and Farshchian, 2018). Moreover, applying the Domestication Theory to technology brings a social element to the evaluation. It provides space to explore the acceptance of the technology at large and ask questions as to whether people are in the stage of initial excitement and high expectations which may not be met, are they fearful of what it might mean for society and the unintended negative consequences of adoption, or are we at the stage of acceptance where the technology is the norm and accepted part of our everyday lives. In our study we can see participants had high expectations of a robotic walker intervention, but they still feared that it would not be able to support meaningful activities and interests in their lives. Negative perceptions about function quality are often mentioned concerning personalized (system-driven) and customizable (user-driven) features of technology (Elers et al., 2018). The common requirements of a cross-cultural sample of older adults for using domestic service robots (Kleanthous et al., 2016) resonates with resources (function quality and ergonomics) and competence (apparent benefits). The cross-cultural, longitudinal study on the Giraff telepresence robot's use showed that the UI's usability and robot's ergonomics were the most influential factors for their acceptance (Orlandini et al., 2016).

Respecting older users' choices and acknowledging their perceptions and life experiences in technology design is critical for acceptance (Kachouie et al., 2014). The sociotechnical perspective on technology design advocates for involving older users not as problems to be fixed but emphasizing their knowledge and skills (Giaccardi et al., 2016; Manzini, 2015). Consequently, older adults show a disposition towards technology that supports doing preferred activities (Mitzner et al., 2010; Vassli and Farshchian, 2018; Yuan et al., 2018). The robotic walker is a critical part of our intervention but it may evoke stigma which raises users fears about the technology. Older adults often avoid using a walker or similar assistive devices because of the fear that they will be labeled as old or disabled, showing other people something is wrong with them or feeling they are letting themselves down (Zwijsen et al., 2011). The most preferred well-being scenarios by healthy participants (Table 1 and 2) show that walker technologies are not stigmatic per se, but may provide opportunities for doing activities and positive user experience. The prevalence of agency influencing behavioral intention indicates that assistive technologies can be designed to create and maintain a positive self-image. According to Lee and Coughlin (2015), older adults favor technology that helps them remain independent in other people's eyes (or gives them competence). Next, they pose that such technology should bring forward their knowledge and skills (or being a resource that can fit existing and evolving needs). Finally, they highlight that it should not assume any sign of aging or decline (thereby facilitating resilience).

Our findings complement these results by revealing and systematizing specific, technology-related factors that can encourage older adults as resources for society and to build their competence to ensure that the technology becomes acceptable to individuals and accepted by society.

*7.1.2. Implications for user requirements of attitudes*

Attitudes refer to values older people would expect to gain from technology supporting their activities as experiential (*feeling good*) and instrumental (*doing good*). Accordingly, Lee and Coughlin (2015) emphasize positive emotional experience and practical values older people see (as useful) for successful technology use.

The technology acceptance models' utility (TAM's perceived usefulness and UTAUT's performance expectancy), being a significant predictor of technology use (Macedo, 2017), reflects the instrumental attitudes, facilitating *doing good* for themselves and others.

In addition to helping manage their health, older adults expect assistive technologies to support different aspects of well-being and to increase their pleasure and satisfaction (Kachouie et al., 2014). *Doing good* appears in the Almere model (as perceived enjoyment and social presence in Heerink et al., 2010) concerning expectations from similar technology. For example, older adults like social robots that can participate in pleasurable activities, such as dancing, listening to music, and playing games (Khosla et al., 2017).

When considering assisted living robotic technologies for well-being, a significant predictor of their use was perceived utility in facilitating health self-monitoring and interactions with known people (Cesta et al., 2018), alerting the support network in emergencies (Elers et al., 2018), or meeting users' expectations regarding concrete tasks (Amirabdollahian et al., 2013). In our results, performance expectancy (Vassli and Farshchian, 2018) resonates with *doing good* including socializing characterized by interests, health and age groups, and *perceived control* or self-management of well-being. The preference for volunteering scenarios shows that this activity can be a powerful behavioral driver for older adults from *doing good*.

*7.1.3. Implications for user requirements of norms*

The requirements of perceived norms structure social influence as the factor of technology acceptance (Venkatesh et al., 2003). Lee and Coughlin (2015) aggregate previous research and classify relevant social influencers as family, peers, and community. Similarly, Vassli and Farshchian (2018) describe significant social influence as peers, family, doctors, and service providers. Along with these groups, younger relatives influenced digital gameplay initiation (De Schutter et al., 2015). Next, similar others' influence as a descriptive reference group is aligned with social learning among peers (Müller et al., 2015). Conversely, Waycott et al. (2016) highlight family circumstances (such as providing care for family members) and existing social norms (e.g., attitudes of family members) as potential barriers for older adults' technology use.

We organize the social factors based on the categories and their influence on older adults - significant others having a descriptive role, and similar others (or reference group) playing a normative role. We extend the reference (peer) group as not only based on age or health but shared interests and hobbies independent of age. These findings suggest new trajectories in technology design, such as facilitating intergenerational activities or volunteering.

The salient barriers to using technology for doing their activities correspond with the older adults' reasons for non-use of the technology reported by Waycott et al. (2016) regarding health problems and learning effort. The inconvenience caused by the technology's monetary cost (e.g., professional photo camera or robot) or increased effort to master it corresponds with the results from the integrated model of technology adoption (Lee and Coughlin, 2015) and interview studies (Elers et al., 2018; Mitzner et al., 2010).

*7.2. Design implications*

The results of our study frame a mixture of physical, mental and social attributes of "successful" aging (Vines et al., 2015a) into a more coherent set of dimensions to explore, create and categorize user requirements for a specific inclusive technology for later life.

Our analysis reveals a complex picture of factors influencing older user intention to use active aging technologies to support their desired activities. In this respect, the requirements can act as a set of design goals, and highlight features that can be used in the design to provide more than satisfying needs, but an enjoyable and desirable experience. In addition, barriers can be reduced by highlighting the benefits which can counteract focus on dislikes and costs. This focus on benefits can encourage uses to expend effort in overcoming the barriers. This approach can improve acceptance of older-friendly technologies that provide mental, physical, and social benefits.

*7.2.1. Design for resilience*

The requirements for resources and competence reveal the desired qualities of a technology (Figure 5) that support user agency as the resilience to the specific circumstances of later life caused by aging's natural process. There is a tendency in societies and older people to live independently in their homes and decrease the burden on healthcare (Pu et al., 2019). In that sense, building on these requirements can facilitate self-management of health as part of daily routines, and more importantly, narrow a gap between well-being and health.

Design for resilience emphasizes older users' roles as resources for the community and their competence from learned knowledge and skills from a lifespan perspective. The latter differentiates older adults from other groups. Our results as illustrated by the most chosen scenarios, indicate that the technology that promotes these elements can consider various volunteering and altruistic activities involving multiple stakeholders, from different formal and informal reference groups to solving important societal problems.

*7.2.2. Design for pleasure*

Design for pleasure brings features that enable feeling good and doing good to the forefront of assistive technology development. Mental satisfaction and physical pleasure are major determinants of feeling good (Figure 5). Accordingly, technology for later life should employ aesthetical representations and services that elicit engaging sensory experiences and positive emotional responses.

Our results confirm active aging narratives on embedding physical activity with social and mental activities (Fernández-Mayoralas et al., 2015; Light et al., 2015; Schutzer and Graves, 2004; Vines et al., 2015a). Existing findings address the complexity of these relations partially, for example, by focusing on self-efficacy (Brassington et al., 2002), social support (Litt et al., 2002), or persuasion (Gallagher and Updegraff, 2012). We provide a more comprehensive view of physical activity in older adulthood concerning factors of behavioral intention. The elements are interrelated and should be equally considered - *doing good* highlights socializing and mobility as the primary gains from doing activities while *feeling good* assumes that physical pleasure from doing the activities is entangled with mental satisfaction.

### *7.3. Limitations*

Our sample is relatively small. Therefore, the work presented in this paper does not provide an exclusive and prescriptive set of user requirements for active aging technology but sets the stage for further inquiry inspired by the initial model supported with empirical findings.

The study described in the paper is exclusively based on the perceptions of a sample of English older adults, and it may affect the generalizability of our findings. Also, the fact that the participants were generally healthy may have influenced inclination towards the personas and scenarios representing healthy and active older adults. Cultural differences could arise when applying the model to collect user requirements from other demographic groups - people from other countries may respond differently. Further research into the differences in using the model's constructs for older adults from different countries and cultures is appropriate.

Given our study's early nature, specific elements such as service or information quality did not emerge. The model of desires provides user-centric terminology and means for collecting and organizing user requirements. However, the importance of its factors may vary, depending on the technology. Further investigation is needed to verify and ground the model for different design cases for active aging and compare its effectiveness to other models and approaches.

## 8. Conclusion

We present the model of user requirements for a more inclusive, active aging technology based on the empirical study with older users. The model builds on factors that influence older adults' intention to engage with preferred activities framed as desires. It paints a holistic picture of older user requirements as follows:

- *mental satisfaction* and *physical pleasure* (feeling good) *and socializing*, *volunteering* and *utility* (doing good),
- *family*, *friends* and *professionals* (significant others), as influencing through encouraging use and *similar others* characterized by common interests and health, and mixed age (reference group) as influencing by use of the technology,

- *ergonomics*, *reliability*, *function quality* and *availability* (resources) and *usability*, *mobility* and *benefits* (competence).

The fine-grained user requirements structure can serve as an analytical tool and help designers to discover and explicate (un)successful design paths or features as early as possible. It has two implications. The model can enable a critical reflection on the existing designs and design processes from a motivational viewpoint. In addition, it can guide new designs and transfer reoccurring, successful combinations into design patterns for active aging technology. A promising line of research would be to examine the interplay among different factors of user intention in specific settings of a recommendation technology for leisure activities in older age.

**Acknowledgment**

Our research was supported by the European Union's Horizon 2020 Research and Innovation Programme under Grant Agreement No. 643644 (ACANTO: A CyberphysicAl social NeTwOrk using robot friends). We thank Mark Mushiba for creating the scenarios' animated storyboards, all participants of the study for their help in evaluating the scenarios, and anonymous reviewers for valuable suggestions for improving the paper.

**References**

Ajzen, I. (1991). The theory of planned behavior. Organizational Behavior and Human Decision Processes, 50(2), 179-211.

Amirabdollahian, F., Akker, R., Bedaf, S., Bormann, R., Draper, H., Evers, V., Pérez, J., Gelderblom, G., Ruiz, C., Hewson, D., Hu, N., Koay, K., Kröse, B., Lehmann, H., Mart, P., Michel, H., Prevot-Huille, H., Reiser, U., Saunders, J., Sorell, T., Stienstra, J., Syrdal, D., Walters, M., & Dautenhahn, K. (2013). Assistive technology design and development for acceptable robotics companions for ageing years, Paladyn, Journal of Behavioral Robotics, 4(2), 94-112.

Anderson, N.D., Damianakis, T., Kröger, E., Wagner, L.M., Dawson, D.R., Binns, M.A., Bernstein, S., Caspi, E. & Cook, S.L. (2014). The benefits associated with volunteering among seniors: a critical review and recommendations for future research. Psychological Bulletin, 140(6), 1505.

Barnard, Y., Bradley, M. D., Hodgson, F., & Lloyd, A. D. (2013). Learning to use new technologies by older adults: Perceived difficulties, experimentation behaviour and usability. Computers in Human Behavior, 29(4), 1715-1724.

Beer, J. M., & Takayama, L. (2011). Mobile remote presence systems for older adults: acceptance, benefits, and concerns. In Proceedings of the 6th international conference on Human-robot interaction, ACM, 19-26.

Brassington, G. S., Atienza, A. A., Perczek, R. E., DiLorenzo, T. M., & King, A. C. (2002). Intervention-related cognitive versus social mediators of exercise adherence in the elderly. American journal of preventive medicine, 23(2), 80-86.

Braun, V., & Clarke, V. (2006). Using thematic analysis in psychology. Qualitative research in psychology, 3(2), 77-101.

Carroll, J. M., Convertino, G., Farooq, U., & Rosson, M. B. (2012). The firekeepers: aging considered as a resource. Universal access in the information society, 11(1), 7-15.

Cesta, A., Cortellessa, G., Fracasso, F., Orlandini, A., & Turno, M. (2018). User needs and preferences on AAL systems that support older adults and their carers. Journal of Ambient Intelligence and Smart Environments, 10(1), 49-70.

Czaja, S. J., Charness, N., Fisk, A. D., Hertzog, C., Nair, S. N., Rogers, W. A., & Sharit, J. (2006). Factors predicting the use of technology: findings from the Center for Research and Education on Aging and Technology Enhancement (CREATE). Psychology and aging, 21(2), 333.

Colcombe, S., & Kramer, A. F. (2003). Fitness effects on the cognitive function of older adults: a meta-analytic study. Psychological science, 14(2), 125-130.

Cornwell, E. Y., & Waite, L. J. (2009). Social disconnectedness, perceived isolation, and health among older adults. Journal of Health and Social Behavior, 50(1), 31-48.

Cozza, M., De Angeli, A., & Tonolli, L. (2017). Ubiquitous technologies for older people. Personal and Ubiquitous Computing, 1-13.

Danielsen, A., Olofsen, H., & Bremdal, B. A. (2016). Increasing fall risk awareness using wearables: a fall risk awareness protocol. Journal of biomedical informatics, 63, 184-194.

Davis, F. D. (1989). Perceived usefulness, perceived ease of use, and user acceptance of information technology. MIS quarterly, 319-340.

De Angeli, A., Jovanovic, M., McNeill, A., Coventry, L., (2020). Desires for Active Ageing Technology. International Journal of Human-Computer Studies, 138, 102412.

De Graaf, M. M., Allouch, S. B., & Klamer, T. (2015). Sharing a life with Harvey: Exploring the acceptance of and relationship-building with a social robot. Computers in Human Behavior, 43, 1-14.

De Schutter, B., Brown, J. A., & Vanden Abeele, V. (2015). The domestication of digital games in the lives of older adults. New Media & Society, 17(7), 1170-1186.

Elers, P., Hunter, I., Whiddett, D., Lockhart, C., Guesgen, H., & Singh, A. (2018). User requirements for technology to assist aging in place: qualitative study of older people and their informal support networks. JMIR mHealth and uHealth, 6(6), e10741.

Farage, M. A., Miller, K. W., Ajayi, F., & Hutchins, D. (2012). Design principles to accommodate older adults. Global journal of health science, 4(2), 2.

Fernández-Mayoralas, G., Rojo-Pérez, F., Martínez-Martín, P., Prieto-Flores, M.E., Rodríguez-Blázquez, C., Martín-García, S., Rojo-Abuín, J.M. & Forjaz, M.J. (2015). Active ageing and quality of life: factors associated with participation in leisure activities among institutionalized older adults, with and without dementia. Aging & mental health, 19(11), 1031-1041.

Gallagher, K. M., & Updegraff, J. A. (2012). Health message framing effects on attitudes, intentions, and behavior: a meta-analytic review. Annals of behavioral medicine, 43(1), 101-116.

Giaccardi, E., Kuijer, L., & Neven, L. (2016). Design for resourceful ageing: intervening in the ethics of gerontechnology. In Design Research Society 50th Anniversary Conference, DRS2016.

Gerling, K., Ray, M., & Evans, A. (2017). Designing for Agency and Compassion: Critical Reflections on Technology to Support Physical Activity in Late Life. Paper presented at the 1st GetAMoveOn Annual Symposium, London, UK.

Gillespie, L. D., Robertson, M. C., Gillespie, W. J., Sherrington, C., Gates, S., Clemson, L. M., & Lamb, S. E. (2012). Interventions for preventing falls in older people living in the community. Cochrane Database Syst Rev, 9(11).

Heerink, M., Kröse, B., Evers, V., & Wielinga, B. (2010). Assessing acceptance of assistive social agent technology by older adults: the almere model. International Journal of Social Robotics, 2(4), 361-375.

Kachouie, R., Sedighadeli, S., Khosla, R., & Chu, M. T. (2014). Socially assistive robots in elderly care: a mixed-method systematic literature review. International Journal of Human–Computer Interaction, 30(5), 369-393.

Kim, E. S., & Konrath, S. H. (2016). Volunteering is prospectively associated with health care use among older adults. Social Science & Medicine, 149, 122-129.

Khosla, R., Nguyen, K., & Chu, M. T. (2017). Human robot engagement and acceptability in residential aged care. International Journal of Human–Computer Interaction, 33(6), 510-522.

Kleanthous, S., Christophorou, C., Tsiourti, C., Dantas, C., Wintjens, R., Samaras, G., & Christodoulou, E. (2016). Analysis of elderly users' preferences and expectations on service robot's personality, appearance and interaction. In International Conference on Human Aspects of IT for the Aged Population. Springer, Cham, 35-44.

Lazar, A., & Nguyen, D. H. (2017). Successful Leisure in Independent Living Communities: Understanding Older Adults' Motivations to Engage in Leisure Activities. In Proceedings of the 2017 CHI Conference on Human Factors in Computing Systems. ACM, 7042-7056.

Lee, C., & Coughlin, J. F. (2015). Perspective: older adults' adoption of technology: an integrated approach to identifying determinants and barriers. Journal of Product Innovation Management, 32(5), 747-759.

Light, A., Leong, T. W., & Robertson, T. (2015). Ageing well with CSCW. In ECSCW 2015: Proceedings of the 14th European Conference on Computer Supported Cooperative Work. Springer International Publishing, 295-304.

Light, A., Pedell, S., Robertson, T., Waycott, J., Bell, J., Durick, J., & Leong, T. W. (2016). What's Special about Aging. Interactions, 23(2), 66-69.

Litt, M. D., Kleppinger, A., & Judge, J. O. (2002). Initiation and maintenance of exercise behavior in older women: predictors from the social learning model. Journal of behavioral medicine, 25(1), 83-97.

Macedo, I. M. (2017). Predicting the acceptance and use of information and communication technology by older adults: An empirical examination of the revised UTAUT2. Computers in Human Behavior, 75, 935-948.

Manzini, E. (2015). Design, when everybody designs: An introduction to design for social innovation. MIT press.

McGlynn, S. A., Kemple, S., Mitzner, T. L., King, C. H. A., & Rogers, W. A. (2017). Understanding the potential of PARO for healthy older adults. International Journal of Human-Computer Studies, 100, 33-47.

McNeill, A., et. al., (2017). User requirements refinement report, ACANTO project Deliverable 1.7. http://www.ict-acanto.eu/wp-content/uploads/2015/04/D1.7.pdf Accessed 17.01.2021.

Mitzner, T.L., Boron, J.B., Fausset, C.B., Adams, A.E., Charness, N., Czaja, S.J., Dijkstra, K., Fisk, A.D., Rogers, W.A. & Sharit, J. (2010). Older adults talk technology: Technology usage and attitudes. Computers in Human Behavior, 26(6), 1710-1721.

Montano, D. E., & Kasprzyk, D. (2008). Theory of reasoned action, theory of planned behavior, and the integrated behavioral model. In K. Glanz, B. K. Rimer & K. Viswanath (Eds.), Health behavior and health education: Theory, Research and Practice, 67-96.

Müller, C., Hornung, D., Hamm, T., & Wulf, V. (2015). Measures and tools for supporting ICT appropriation by elderly and non tech-savvy persons in a long-term perspective. In ECSCW 2015: Proceedings of the 14th European Conference on Computer Supported Cooperative Work, Springer International Publishing.

Nomis. (2017). Official Labour Market Statistics. https://www.nomisweb.co.uk/ Accessed 17.01.2021.

Orlandini, A., Kristoffersson, A., Almquist, L., et al., (2016). Excite project: A review of forty-two months of robotic telepresence technology evolution. Presence: Teleoperators and Virtual Environments, 25(3), 204-221.

Parra, C., Silveira, P., Far, I.K., Daniel, F., De Bruin, E.D., Cernuzzi, L., D'Andrea, V. & Casati, F. (2014). Information technology for active ageing: A review of theory and practice. Foundations and Trends® in Human–Computer Interaction, 7(4), 351-448.

Peruzzini, M., & Germani, M. (2014). Designing a user-centred ICT platform for active aging. In 2014 IEEE/ASME 10th International Conference on Mechatronic and Embedded Systems and Applications (MESA), 1-6.

Preedy, V. R. (2010). Handbook of disease burdens and quality of life measures, 4271-4272. In R. R. Watson (Ed.). New York: Springer.

Preece, J., Rogers, Y., & Sharp, H. (2015). Interaction design: beyond human-computer interaction. John Wiley & Sons.

Pu, L., Moyle, W., Jones, C., & Todorovic, M. (2019). The effectiveness of social robots for older adults: a systematic review and meta–analysis of randomized controlled studies. The Gerontologist, *59*(1), e37-e51.

Robinson, H., MacDonald, B., & Broadbent, E. (2014). The role of healthcare robots for older people at home: a review. International Journal of Social Robotics, 6(4), 575-591.

Schutzer, K. A., & Graves, B. S. (2004). Barriers and motivations to exercise in older adults. Preventive medicine, 39(5), 1056-1061.

Silverstone, R., & Haddon, L. (1996). Design and the domestication of information and communication technologies: Technical change and everyday life. In: Silverstone R and Mansell R (eds), Communication by design: The politics of information and communication technologies, Oxford University Press.

Spaniolas, K., Cheng, J. D., Gestring, M. L., Sangosanya, A., Stassen, N. A., & Bankey, P. E. (2010). Ground level falls are associated with significant mortality in elderly patients. Journal of Trauma and Acute Care Surgery, 69(4), 821-825.

Strawbridge, W. J., Deleger, S., Roberts, R. E., & Kaplan, G. A. (2002). Physical activity reduces the risk of subsequent depression for older adults. American journal of epidemiology, 156(4), 328-334.

Stel, V. S., Smit, J. H., Pluijm, S. M., & Lips, P. (2004). Consequences of falling in older men and women and risk factors for health service use and functional decline. Age and ageing, 33(1), 58-65.

Stenner, P., McFarquhar, T., & Bowling, A. (2011). Older people and 'active ageing': Subjective aspects of ageing actively. Journal of health psychology, 16(3), 467-477.

Sutcliffe, A. (2003). Scenario-based requirements engineering. In Proceedings of the11th IEEE international Requirements Engineering Conference, 320-329.

Sun, F., Norman, I. J., & While, A. E. (2013). Physical activity in older people: a systematic review. BMC public health, 13(1), 449.

Suwa, S., Tsujimura, M., Ide, H., Kodate, N., Ishimaru, M., Shimamura, A., & Yu, W. (2020). Home-care Professionals' Ethical Perceptions of the Development and Use of Home-care Robots for Older Adults in Japan. International Journal of Human–Computer Interaction, 36(14), 1-9.

Tay, L., & Diener, E. (2011). Needs and subjective well-being around the world. Journal of personality and social psychology, 101(2), 354.

The European Union's Horizon 2020 Research and Innovation Programme - Societal Challenge 1 (DG CONNECT/H). Grant agreement 643644. "ACANTO - A CyberphysicAl social NeTwOrk using robot friends". http://www.ict-acanto.eu/ Accessed 17.01.2021.

Tinetti, M. E., & Williams, C. S. (1998). The effect of falls and fall injuries on functioning in community-dwelling older persons. The Journals of Gerontology Series A: Biological Sciences and Medical Sciences, 53(2), 112-119.

Truong, K. N., Hayes, G. R., & Abowd, G. D. (2006). Storyboarding: an empirical determination of best practices and effective guidelines. In Proceedings of the 6th AMCM conference on Designing Interactive systems, 12-21.

Uzor, S., & Baillie, L. (2014). Investigating the long-term use of exergames in the home with elderly fallers. In Proceedings of the 32nd annual ACM conference on Human factors in computing systems, 2813-2822.

Varma, V.R., Tan, E.J., Gross, A.L., Harris, G., Romani, W., Fried, L.P., Rebok, G.W. & Carlson, M.C. (2016). Effect of community volunteering on physical activity: A randomized controlled trial. American journal of preventive medicine, 50(1), 106-110.

Vandemeulebroucke, T., de Casterlé, B. D., & Gastmans, C. (2018). How do older adults experience and perceive socially assistive robots in aged care: a systematic review of qualitative evidence. Aging & mental health, 22(2), 149-167.

Vassli, L. T., & Farshchian, B. A. (2018). Acceptance of Health-Related ICT among Elderly People Living in the Community: A Systematic Review of Qualitative Evidence. International Journal of Human–Computer Interaction, 34(2), 99-116.

Venkatesh, V., Morris, M. G., Davis, G. B., & Davis, F. D. (2003). User acceptance of information technology: Toward a unified view. MIS quarterly, 425-478.

Vines, J., Pritchard, G., Wright, P., Olivier, P., & Brittain, K. (2015a). An age-old problem: Examining the discourses of ageing in HCI and strategies for future research. ACM Transactions on Computer-Human Interaction (TOCHI), 22(1), 2.

Vines, J., Wright, P. C., Silver, D., Winchcombe, M., & Olivier, P. (2015b). Authenticity, Relatability and Collaborative Approaches to Sharing Knowledge about Assistive Living Technology. In Proceedings of the 18th ACM Conference on Computer Supported Cooperative Work & Social Computing, 82-94.

Yuan, C. W., Hanrahan, B. V., Rosson, M. B., & Carroll, J. M. (2018). Coming of Old Age: Understanding Older Adults' Engagement and Needs in Coproduction Activities for Healthy Ageing. Behaviour & Information Technology, 1-15.

Yusif, S., Soar, J., & Hafeez-Baig, A. (2016). Older people, assistive technologies, and the barriers to adoption: A systematic review. International journal of medical informatics, 94, 112-116.

Waycott, J., Vetere, F., Pedell, S., Morgans, A., Ozanne, E., & Kulik, L. (2016). Not For Me: Older Adults Choosing Not to Participate in a Social Isolation Intervention. In Proceedings of the 2016 ACM CHI Conference on Human Factors in Computing Systems, 745-757.

Well-being for life. (2017). Know your city: a profile of the people living in Newcastle: The people living, working or learning in Newcastle https://bit.ly/397DCQE Accessed 17.01.2021.

Wilkinson, C. R., & De Angeli, A. (2014). Applying user centred and participatory design approaches to commercial product development. Design Studies, 35(6), 614-631.

Wilson, R. S., Hebert, L. E., Scherr, P. A., Dong, X., Leurgens, S. E., & Evans, D. A. (2012). Cognitive decline after hospitalization in a community population of older persons. Neurology, 78(13), 950-956.

Whelan, S., Murphy, K., Barrett, E., Krusche, C., Santorelli, A., & Casey, D. (2018). Factors affecting the acceptability of social robots by older adults including people with dementia or cognitive impairment: A literature review. International Journal of Social Robotics, 1-26.

World Health Organization - WHO. (2002). Active geing: a policy framework. https://bit.ly/392oEuZ (p. 4) Accessed 17.01.2021.

Wu, Y. H., Wrobel, J., Cornuet, M., Kerhervé, H., Damnée, S., & Rigaud, A. S. (2014). Acceptance of an assistive robot in older adults: a mixed-method study of human-robot interaction over a 1-month period in the Living Lab setting. Clin Interv Aging, 9, 801-811.

Zwijsen, S. A., Niemeijer, A. R., & Hertogh, C. M. (2011). Ethics of using assistive technology in the care for community-dwelling elderly people: an overview of the literature. Aging & mental health, 15(4), 419-427.

**Appendix**

The interview script is initially reported in the ACANTO project deliverable (McNeill et al., 2017). The starting questions concerned facilitators and barriers, whereas the remaining questions examined desires (see Figure 2) as factors influencing behavioral intentions to use the ACANTO system.

| Category | Question |
|---|---|
| Facilitators | What abilities does this person have that the system can encourage? |
| Barriers | What problems does this person face ? How well do you think the system solves these problems? |
| Feeling good | Why would this person use this system? How would the system make the person feel? |
| Doing good | What might the person gain from using the system? |
| Significant others | How would other people's views affect the user's willingness to use the system? |
| Reference group | How would it affect the person's willingness to use the system if other people used it? |
| Resources | What sort of things might help the person to use the system? What sort of things might make it hard for the person to use the system? |
| Competence | How important is it that the person finds it easy to use the system? |
| Counter question | Why would this person not use this system? |

**About the Authors**

**Dr Mlađan Jovanović** received PhD in Computer Science at the University of Belgrade in 2013. He is an Assistant Professor at Singidunum University, Belgrade. Before academia, he worked in the industry, focusing on the interactive computing systems. His main research interests include technologies for later life and human-centered AI.

**Dr Antonella De Angeli** is a Full Professor of Human-Computer Interaction at the Free University of Bolzano, Italy. She holds a PhD in Experimental Psychology from the University of Trieste. Antonella has a track record of successful social innovation projects. Recently, her research interests include active aging and participatory design.

**Dr Andrew McNeill** is a senior lecturer in Psychology at Northumbria University, Newcastle, UK. He regularly conducts research in both social psychology and human-computer interaction, especially in privacy and technologies for well-being of older adults.

**Dr Lynne Coventry** is the Director of PaCT Lab (Psychology and Communication Technology) at the University of Northumbria, UK. She received a PhD in Psychology. She has been working on the research that incorporates understanding of people, their use and acceptance of technology into the requirements and design process.